\documentclass[%
 reprint,
superscriptaddress,
 amsmath,amssymb,aps,
 prb,
]{revtex4-2}

\usepackage{graphicx}
\usepackage{dcolumn}
\usepackage{bm}
\usepackage{xcolor}
\usepackage{amsmath,amssymb}
\usepackage{float}
\usepackage{mathtools}
\usepackage{soul}

\usepackage{xcolor}
\newcommand\shorthandon{\catcode`@=\active }
\newcommand\shorthandoff{\catcode`@=12 }

\shorthandon
\def@#1@ {\textcolor{red}{#1} }%
\shorthandoff

\begin{document}
\shorthandon

\title{Universal exciton polariton logic gates in Ouroboros rings}

\author{Tobias Schneider}
\affiliation{Department of Physics and Center for Optoelectronics and Photonics Paderborn (CeOPP), Paderborn University, 33098 Paderborn, Germany}

\author{Stefan Schumacher}
\affiliation{Department of Physics and Center for Optoelectronics and Photonics Paderborn (CeOPP), Paderborn University, 33098 Paderborn, Germany}
\affiliation{Institute for Photonic Quantum Systems (PhoQS), Paderborn University, 33098 Paderborn, Germany}
\affiliation{Wyant College of Optical Sciences, University of Arizona, Tucson, Arizona 85721, USA}

\author{Xuekai Ma}
\affiliation{Department of Physics and Center for Optoelectronics and Photonics Paderborn (CeOPP), Paderborn University, 33098 Paderborn, Germany}

\begin{abstract}
All-optical logic gates have significantly advanced over a diverse range of photonic systems, boosted by intricate nonlinearities that facilitate the engineering of complex logic operations. Here, we demonstrate that in semiconductor microcavities, polariton condensates trapped in Ouroboros-shaped rings form specifically charged vortices, determined by the strength of nonlinearity and the excitation method. Quantized vortex phases encode binary digits that can be nonresonantly controlled by optical pulses incident directly upon the ring, enabling logic operations. By interconnecting three polariton Ouroboros rings, we realize a universal set of logic gates (AND, OR, NIMPLY) fundamental to functional polaritonic devices. The Ouroboros structures are highly customizable, providing a robust and promising platform for exploring more complex logic operations.
\end{abstract}

\maketitle

\section{Introduction}
Logic gates serve as the fundamental building blocks of digital electronic circuits. A functionally complete set of logic gates is capable of performing all basic Boolean operators (AND, OR, and NOT). All-optical logic gates have generated significant interest across various platforms for the development of all-optical circuits~\cite{singh2014all,salmanpour2015photonic,anagha2022review}. In photonic crystals, for instance, linear interference enables standard logic gates such as AND, OR, and XOR, while nonlinearity supports more robust functions like NAND and XNOR~\cite{salmanpour2015photonic,jot2020all}. Another prominent platform is the semiconductor optical amplifier (SOA), favored for its strong nonlinear effects and high potential for integration~\cite{sobhanan2022semiconductor,cui2026advances}. In these systems, logic gates are implemented by using nonlinearity to manipulate optical signals. 

In semiconductor micro- and nanostructures, photons can also be harnessed via strong light-matter coupling. In this regime, exciton-polaritons (hereafter polaritons) can form as quasiparticles composed of both excitons and photons. Due to their excitonic nature, polariton-polariton interactions lead to significant nonlinearity, enabling intriguing nonlinear phenomena such as bistability~\cite{PhysRevA.69.023809,PhysRevLett.101.266402,PhysRevLett.112.076402,PhysRevLett.120.225301}, solitons~\cite{PhysRevLett.102.153904,sich2012observation,hivet2012half,PhysRevLett.111.146401,PhysRevLett.118.157401}, and vortices~\cite{lagoudakis2008quantized,roumpos2011single,sanvitto2010persistent,tosi2012geometrically,sala2015spin,ma2020realization}. These phenomena also play a role for both classical and quantum computing based on polaritons~\cite{kavokin2022polariton}, particularly through the manipulation of quantized topological charges of vortices~\cite{PhysRevLett.113.200404,ma2020realization,PhysRevLett.131.136901,alyatkin2024antiferromagnetic,PhysRevResearch.3.013099,barrat2024qubit}. Basic logic gates have been proposed~\cite{PhysRevB.82.033302,PhysRevB.87.195305} and realized~\cite{ballarini2013all}, especially in recent studies of room-temperature polariton condensates in organic~\cite{zasedatelev2019room,sannikov2024room}, ZnO~\cite{li2024all}, and perovskite~\cite{jin2026dynamic} microcavities. This progress, combined with the demonstrations of the polaritons' ultrafast response, establishes polariton systems as an ideal platform for more complex integrated all-optical circuits. 

\begin{figure}[b]
\centering
{\includegraphics[width=1\linewidth]{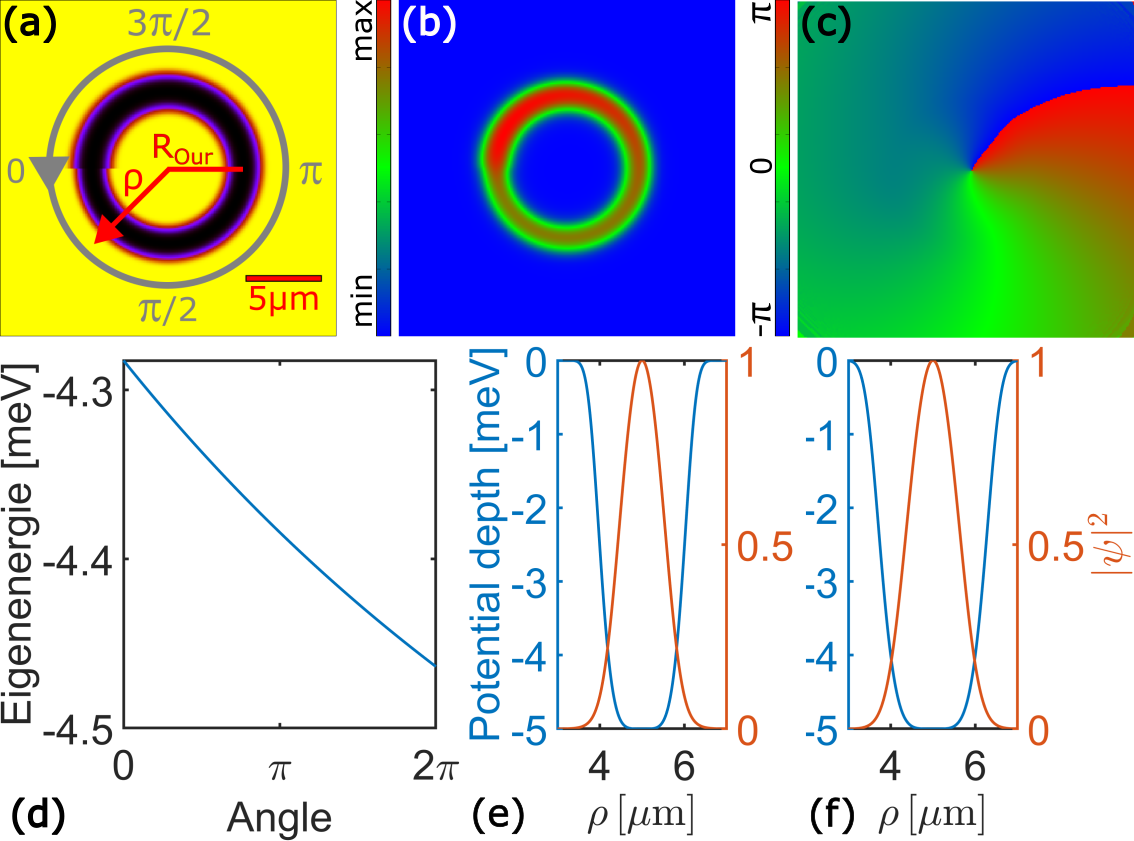}}
\caption{\textbf{Ouroboros ring structure and trapped states.} (a) Profile of an Ouroboros ring with the radius $R_{\textup{Our}}$. The minimum width of the potential is $W_t$ at angle 0 (below the seam), while the maximum width is $W_0$ at angle $2\pi$ (above the seam). $\rho$ indicates the radial direction. (b) Density ($|\psi|^2$, in $\mu\textup{m}^{-2}$) profile of the trapped condensate with counter-clockwise rotation ($m=+1$), as can be seen from the phase in (c). (d) Angle-dependent eigenenergy of the ground state of the 1D potential well in the radial direction. Example 1D potential distributions (blue lines) and the ground states (orange lines) at the angle (e) 0 and (f) $2\pi$.}
\label{fig:1}
\end{figure}

While basic logic gates have been realized in polariton systems, complex logic gates, such as implication (IMPLY) and non-implication (NIMPLY), remain a challenge, as they typically require interconnecting multiple basic gates, necessitating specific architectural designs. In the present work, we demonstrate an Ouroboros ring-shaped potential that supports distinct stable polariton currents, which can be manipulated by nonlinearity and incoherent optical beams. Nonresonantly controllable polariton currents facilitate the implementation of not only the AND and OR gate, but also the NIMPLY gate, which is an inhibitor gate for conditional logic and has advanced applications in, for example, synthetic biology~\cite{siuti2013synthetic,jung2022programming}. This complex NIMPLY gate has been recently demonstrated in metamaterials for optical information processing~\cite{jiang2024photonic}. Importantly, these three gates constitute a functionally complete set for logic operations. Unlike standard rings~\cite{PhysRevB.100.245304,barkhausen2020multistable}, the Ouroboros geometry favors specific polariton currents, simplifying logic gate construction. Furthermore, these rings allow for polariton leakage at the junction, enabling efficient coupling of multiple rings for the engineering of even more complex logic circuits.

\textit{Theoretical model.} -- The dynamics of polariton condensates in semiconductor microcavities is governed by an extended Gross-Pitaevskii (GP) equation coupled to the evolution equation of an excitation reservoir that is nonresonantly excited by an incoherent pump~\cite{PhysRevLett.99.140402}, i.e.,
\begin{equation}
\label{GP_psi}
\begin{split}
i\hbar \frac{\partial \psi(\mathbf{r},t)}{\partial t} &= 
\Big[-\frac{\hbar^2}{2m_{\textup{eff}}}\nabla^2 - i\hbar \frac{\gamma_c}{2} + g_c |\psi(\mathbf{r},t)|^2 \\
&\quad + \big(i\hbar \frac{R}{2} + g_r \big) n_R + V(\mathbf{r}) \Big] \psi(\mathbf{r},t)
\end{split}
\end{equation}
\begin{equation}
\begin{aligned}
\label{GP_n}
    \frac{\partial n_R(\textbf{r},t)}{\partial t} = -(\gamma_r + R |\psi(\textbf{r},t)|^2) n_R(\textbf{r},t) + P(\textbf{r},t).
\end{aligned}
\end{equation}
Here, $\psi(\textbf{r},t)$ is the wavefunction of the polariton condensate and $n_R(\textbf{r},t)$ is the density of the excitation reservoir. The effective mass is $m_{\textup{eff}}=10^{-4} m_e$ with $m_e$ being the free electron mass. The decay rate of the condensate is $\gamma_c=0.2\ \textup{ps}^{-1}$, and the nonlinearity strength is described by $g_\textup{c}=6\ \mu\textup{eV}\ \mu\textup{m}^2$. The condensate is fed from the reservoir with a condensation rate $R=0.01\ \textup{ps}^{-1}\ \mu\textup{m}^2$, and the interaction with the reservoir is $g_\textup{r}=12\ \mu\textup{eV}\ \mu\textup{m}^2$. The reservoir also decays with $\gamma_\textup{r}=0.3\ \textup{ps}^{-1}$ and is sustained by the incoherent pump $P(\textbf{r},t)$. In this work, the potential $V(\textbf{r})$ takes the shape of a ring with its width smoothly varying in the azimuthal direction, forming an Ouroboros-like structure as shown in Fig.~\ref{fig:1}(a). This potential configuration (mathematical definition in the SM) contains an intersection at the point where the potential width jumps. This intersection is henceforth referred to as the seam. Here, we set the width of the broader side of the seam to $W_0$ and the narrower side to $W_t$.

\section{Polariton currents in Ouroboros rings}
In this potential configuration, although the potential depth remains uniform at 5 meV, the angle-dependent potential width induces varying local ground-state energies [Fig.~\ref{fig:1}(d-f)]. This creates an energy gradient that favors a current flowing in the direction of increasing width, i.e., counter-clockwise rotation. Consequently, as shown in Fig.~\ref{fig:1}(b,c), a vortex with topological charge $m=+1$ is formed under excitation with a ring-shaped pump centered on the potential ring. Detailed descriptions of the pump configuration and the excitation schematic are available in the SM. The default potential parameters are $R_\textup{Our}=5\ \mu$m, $W_0=2.7\ \mu$m, and $W_t/W_0=0.8$. The pump intensity is $P_0 = 227\ \textup{eV}\ \mu\textup{m}^2$ and the width is $1\ \mu$m. 

\begin{figure}[t]
\centering
{\includegraphics[width=1\linewidth]{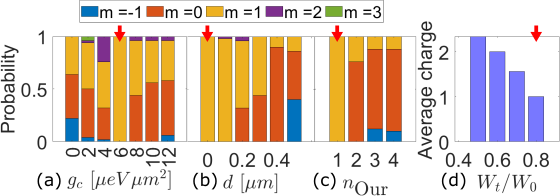}}
\caption{\textbf{Probability of differently charged vortices.} Influence of (a) the strength of nonlinearity $g_\textup{c}$, (b) the pump offset $d$, and (c) the seam count (periods) $n_\textup{Our}$ on the observation probability of different vortices. For each case, 50 calculations were performed with unique initial noise. (d) Dependence of the average charge of 50 calculations on the ratio of $W_t/W_0$. Red arrows indicate the default case.}
\label{fig:2}
\end{figure}

In addition to the $m=+1$ state, $m=-1$ and $m=0$ vortices are also the eigenstates with slightly higher energies. In the default case, see the red arrow in Fig.~\ref{fig:2}(a), all excitations consistently result in $m=+1$ charged vortices. To ensure statistical robustness, we repeated each calculation 50 times using the numerical solver PHOENIX~\cite{wingenbach2025phoenix}, with each run initialized with unique random noise. As the nonlinear effect is adjusted, which can also be achieved by keeping the strength of the nonlinearity fixed and varying the pump intensity, the probability of observing the $m=+1$ vortex decreases, while the $m=0$ state becomes more prevalent. In contrast, the $m=-1$ vortex rarely forms, since its flow opposes the established energy gradient. In certain cases, higher-order vortices were also observed. 

\begin{figure*}[t]
\centering
{\includegraphics[width=1\linewidth]{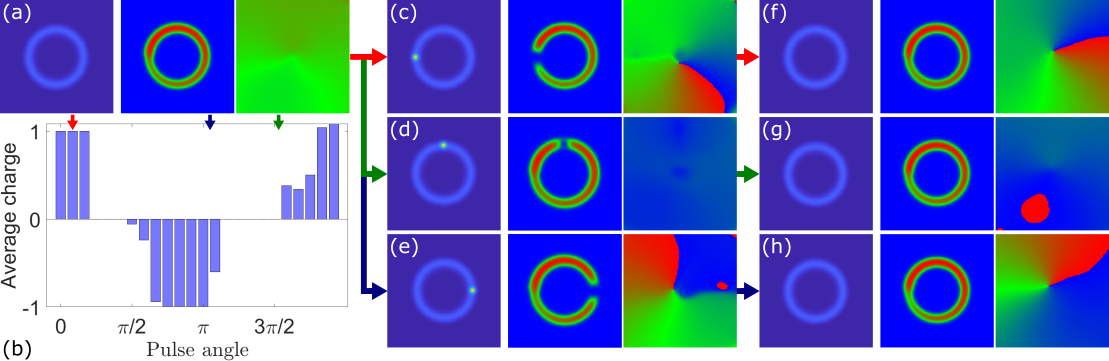}}
\caption{\textbf{Vortex switching.} (a) An initial $m=0$ state (left: pump, middle: density, right: phase). (b) Dependence of the average charge (50 calculations) on the angular position of the control pulse on the ring. (c-e) States when the control pulse is applied at the angle (c) 0, (d) $\pi$, and (e) $3\pi/2$, respectively indicated by the arrows in (b). (f-h) Final steady states of (c-e), respectively, with (f) $m=+1$, (g) $m=0$, and (h) $m=-1$ in the absence of the pulses.}
\label{fig:3}
\end{figure*}

Another critical factor determining the topological charge is the ring-pump offset $d$ relative to the Ouroboros ring. In the default case without offset, the potential ring undergoes homogeneous azimuthal excitation because the ring pump is perfectly centered on the potential. In this case, the seam behaves analogously to a p-n junction, with the condensate flowing unidirectionally from the broader side to the narrower side. To reverse the current flow, we incrementally shift the ring pump toward the bottom-left direction [indicated by the red arrow in Fig.~\ref{fig:1}(a); see also SM] to enhance the polariton density within the narrower part of the Ouroboros ring. A smaller offset, $d=0.2\ \mu$m for instance, reduces the probability of the $m=+1$ state and allows for $m=0$ vortices to emerge [Fig.~\ref{fig:2}(b)]. Increasing the offset to $d=0.4\ \mu$m significantly suppresses the $m=+1$ state, leaving $m=0$ dominant. The $m=-1$ vortex is activated at $d=0.5\ \mu$m, reaching a formation probability ($\sim$40\%) comparable to that of $m=0$. Further increasing the offset to $d>0.5\ \mu$m reduces the spatial overlap between the pump and the potential ring, preventing the formation of stable states.

Increasing the number of seams or periods within the ring reduces the length of each individual period, which favors the $m=0$ state over others [Fig.~\ref{fig:2}(c)]. This occurs because the shortened path length provides insufficient distance for a stable current to develop. Moreover, the reflection of the condensate from the broader side of the seams intensifies as the seam count increases. The current flowing dynamics are also strongly influenced by the azimuthal gradient of the potential width, represented by a ratio $W_t/W_0$ where smaller values indicate steeper gradients. In the default configuration with $W_t/W_0=0.8$, the width difference between the two sides is rather small. Decreasing this ratio increases the slope of the angle-dependent ground-state energy, cf. Fig.~\ref{fig:1}(d), which accelerates the condensate flow and triggers higher-order vortices with $m>1$. Figure~\ref{fig:2}(c) shows the average charge over 50 calculations at specific ratios. For the default case, this value is 1, indicating the exclusive presence of the $m=+1$ state. Smaller ratios result in the average change exceeding 1.5, confirming the excitation of higher-order modes. These higher-order vortices appear more frequently at the ratio 0.5. Decreasing this ratio to $W_t/W_0<0.5$ constricts the flow of the condensate into petal-like patterns. Notably, these vortex states in Ouroboros rings are stable against random potential disorders of up to $\pm0.8$~meV, with a correlation length of $\sim2-3\ \mu$m (same below).

\section{Nonresonant control of polariton currents}
Although vortices formed under the default condition with initial noise typically retain a topological charge of $m=+1$, the $m=0$ and $m=-1$ states are also accessible due to nonlinearity and specific initial conditions. To enhance the accessibility of these states, we increase the nonlinearity to $12\ \mu\textup{eV}\ \mu\textup{m}^2$ and set the pump offset to $d = 0.1\ \mu\textup{m}$. These three stable states can be switched using a Gaussian-shaped incoherent pulse (see the definition in SM) with a spatial extent comparable to the potential width. 

Figure~\ref{fig:3}(b) illustrates the dependence of the average charge on the angular position of the applied pulse. At the angle 0, the $m=+1$ state remains the sole outcome. Shifting the angle to $\pi$ (see the blue arrow), however, isolates the $m=-1$ state. The $m=0$ state dominates when the average charge approaches 0 (around angles $\pi/2$ and $3\pi/2$). For example, at angle $3\pi/2$ (see the green arrow), the average charge is 0, and neither the $m=+1$ nor the $m=-1$ states form. Examples of the switching dynamics are shown in Fig.~\ref{fig:3}, beginning from a $m=0$ state [Fig.~\ref{fig:3}(a)]. When a control pulse is applied at angle 0 [Fig.~\ref{fig:3}(c)], it acts as a barrier~\cite{ma2020realization,PhysRevResearch.3.013099,barrat2024qubit,voronova2025exciton}, creating a density gap that forces the phase into a counter-clockwise spiral. Consequently, the extinction of the pulse leads to a stable $m=+1$ state [Fig.~\ref{fig:3}(f)]. If the pulse is applied at an angle $3\pi/2$ [Fig.~\ref{fig:3}(d)], the result remains unchanged. At angle $\pi$, the pulse switches the state to $m=-1$ [Figs.~\ref{fig:3}(e) and ~\ref{fig:3}(h)]. The switching dynamics are robust and independent of the initial state. In each process, the pulse is applied for approximately $24$~ps at an intensity five times greater than that of the ring pump. This switching process is also robust against random potential disorders of up to $\pm 0.8$ meV.

\begin{figure*}[t]
\centering
{\includegraphics[width=1\linewidth]{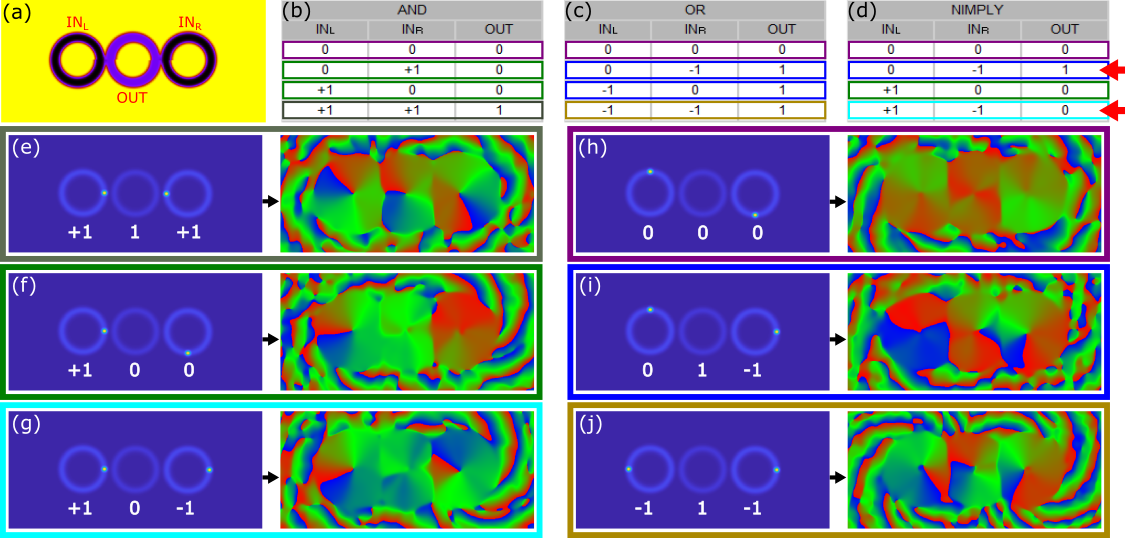}}
\caption{\textbf{Logic gate implementation.} (a) Potential structure for logic gates. Left and right rings are defined as input $IN_L$ and $IN_R$, respectively, and the output ring $OUT$ is placed in between. Tables of the logic operation (b) AND, (c) OR, and (d) NIMPLY. The same operations, including the rotationally symmetric ones, are marked by the same colored frames. (e-j) Realization of the operations in (b-d) with the input and output signals indicated in the excitation schemes (left panels). Right panels are the phases of the corresponding final states. The same coloured frames in (b-d) and (e-j) correspond to each other.}
\label{fig:4}
\end{figure*}

\section{Logic gates}
The realized switching dynamics provide a robust framework for implementing all-optical logic gates in Ouroboros ring architectures. A key advantage of the Ouroboros geometry is its seam, which facilitates inter-ring connectivity more effectively than standard ring potentials. In this work, we propose a linear arrangement of three Ouroboros rings, where the central ring incorporates two periods to optimize its coupling with the flanking input rings, see Fig.~\ref{fig:4}(a). While the above studies focus on single-period geometries, increasing the number of periods offers an additional degree of freedom for tailoring polariton currents. In the structure in Fig.~\ref{fig:4}(a), the two rings on both sides are used as inputs, while the ring in the middle is used as an output. Since there are two repetitions in the middle ring, we employ a shallower potential depth of 3.5 meV to adapt the energy levels of the single-period neighbours and to increase the stability of the input rings. Their contact allows efficient polariton tunnelling, enhancing the inter-ring coupling. The system is driven by three simultaneous ring-shaped pumps. Logic operations are executed by applying control pulses to the input rings to manipulate their internal currents and consequently dictate the resulting current state of the output ring. 

Due to the rotational symmetry of the structure and the enhanced coupling of the rings, this configuration allows for three distinct logic gates: AND, OR, and NIMPLY, as introduced in Figs.~\ref{fig:4}(b-d). Here, we define binary 0 as the topological charge $m=0$, i.e., without currents, and binary 1 as the topological charges $m=+1$ or $m=-1$, i.e., carrying currents. These three logic gates are categorized into six operational states, distinguished by coloured frames in Figs.~\ref{fig:4}(b-d). For example, the ($0$$|$$0$$\rightarrow$$0$) operation is fundamental to all gates, and its implementation is illustrated in Fig.~\ref{fig:4}(h). Based on the result in Fig.~\ref{fig:3}(b), a control pulse applied at angle $\pi/2$ ensures the $m=0$ state. Thus, two pulses applied to both input rings yield a stable 0 output [Fig.~\ref{fig:4}(h)]. Note that since the left input ring is rotated 180 degrees relative to the right one, the control pulse angles are adjusted accordingly. There are two possible configurations if only one of the inputs has a current: If the input is $m=+1$, the output is $m=0$ (binary 0); If the input is $m=-1$, the output is $m=+1$  (binary 1). The former (latter) is the basic logic for the AND (OR) and NIMPLY gate, see the green (blue) frames in Fig.~\ref{fig:4}(b-d). The realization of the ($+1$$|$$0$$\rightarrow$$0$) [the same as ($0$$|$$+1$$\rightarrow$$0$)] operation is presented in Fig.~\ref{fig:4}(f). In this case, a counter-clockwise current in the left ring attempts to trigger a clockwise current in the central ring via tunnelling. However, because the counter-clockwise current is the central ring's preferred state, these competing influences force the output into the $m=0$ state. Conversely, if the left current is clockwise, it directly triggers the central ring's preferred counter-clockwise rotation, switching the output to $m=+1$ even though the right ring remains $m=0$, that is, the operation ($-1$$|$$0$$\rightarrow$$1$) [the same as the operation ($0$$|$$-1$$\rightarrow$$1$) in Fig.~\ref{fig:4}(i)]. 

As illustrated in Figure~\ref{fig:4}(b-d), each gate also includes a unique operation, which can be summarized as: If both inputs have the same charge, the output binary is 1; If the inputs have opposite charges, the output binary is 0. The operation ($-1$$|$$-1$$\rightarrow$$1$) is particularly intuitive, as clockwise currents from both the left and right ring constructively trigger the preferred counter-clockwise current in the central ring [Fig.~\ref{fig:4}(j)]. In contrast, when the inputs have opposite signs, see Fig.~\ref{fig:4}(g), their competing influences cancel each other out, leaving the central condensate in a quiescent $m=0$ state. The unintuitive operation is ($+1$$|$$+1$$\rightarrow$$1$) realised in Fig.~\ref{fig:4}(e), in which $\pi$-phase jumps occur at the contact points between the three rings and the phase locking enables the counter-clockwise current in the central ring. Therefore, three distinct logic gates are successfully implemented within the Ouroboros rings under the control of only two incoherent pulses. Note that due to the reduced potential depth of the output ring, the robustness of the gates against random disorder is also reduced to around $\pm 0.4$ meV. Although strong sample disorder or pump fluctuation can impact stability, the Ouroboros geometry is highly adaptable, allowing for structural modifications to achieve an optimal default configuration.

It is worth noting that the NIMPLY gate can also function as a NOT gate by utilizing the two specific operations highlighted by the red arrows in Fig.~\ref{fig:4}(d). Integrating this NOT functionality with either the AND or OR gates allows for the realization of NAND and NOR gates, respectively. Since both NAND and NOR gates are functionally complete, any logical operation can be implemented by combining multiple such gates. The Ouroboros-based gate architecture establishes a robust platform for universal binary logic driven by condensate vortices.

\section{Conclusion}
To conclude, we have demonstrated a functionally complete set of logic gates (AND, OR, and NIMPLY) in polariton Ouroboros rings. The unique potential gradient in these rings supports multiple states for polariton currents, with a geometry-induced preference for a specific circulation direction. This preferred current direction is sensitive to both nonlinearity and the excitation method, and the direction of its flow can be actively switched using short pulses that are incoherent with the condensate, potentially also enabling electric control in specific realizations~\cite{PhysRevLett.131.136901}. The Ouroboros geometry facilitates strong inter-ring coupling, paving the way for the engineering of complex logic circuits and advanced polariton-based devices.

\section*{Acknowledgements}
This work was supported by the Deutsche Forschungsgemeinschaft (DFG) (No. 519608013 and No. 467358803) and the Paderborn Center for Parallel Computing, PC$^2$.

%

\end{document}